\documentclass[prd,aps,floatfix,nofootinbib,11 pt]{revtex4}
\usepackage{amsmath,graphicx,color,epsfig}

\input{tcilatex}

\begin{document}

\title{Distinguishing quantum channels via magic squares game}
\author{M. Ramzan\thanks{%
mramzan@phys.qau.edu.pk} and M. K. Khan}

\address{Department of Physics Quaid-i-Azam University \\
Islamabad 45320, Pakistan}

\date{\today}

\begin{abstract}
We study the effect of quantum memory in magic squares game when
played in quantum domain. We consider different noisy quantum
channels and analyze their influence on the magic squares quantum
pseudo-telepathy game. We show that the probability of success can
be used to distinguish the quantum channels. It is seen that the
mean success probability decreases with increase of quantum noise.
Where as the mean success probability increases with increase of
quantum memory. It is also seen that the behaviour of amplitude
damping and phase damping channels is similar. On the other hand,
the behaviour of depolarizing channel is similar to the flipping
channels. Therefore, the probability of success of the game can be
used to distinguish the quantum channels.\newline
\end{abstract}

\maketitle

\address{Department of Physics Quaid-i-Azam University \\
Islamabad 45320, Pakistan}

\newline

\vspace*{1.0cm}

\vspace*{1.0cm}



\section{Introduction}

Applications of game theory \cite{RasE} can be found in several research
areas, such as economics, biology, physics and computer sciences. In the
recent past, rapid interest has been developed in the discipline of quantum
information \cite{NieMA} that has led to the creation of quantum game theory
\cite{MeyDA}. During last few years, number of authors have contributed to
the development of quantum game theory [4-9]. In this direction, much work
has been done on quantum prisoners' dilemma game [10-12] and other games
have been converted to the quantum realm including the battle of the sexes
[8, 13], the Monty Hall problem [14, 15], the rock-scissors-paper [16], and
the others [17-20]. Quantum pseudo-telepathy game [21] is something which
can not be won in the classical world without communication but can be won
in the quantum world using entangled state without any use of classical
communication. Brassard et al. [21] have shown that how to win the magic
square game for $n=3$ with certainty by sharing a two qubit entanglement
between Alice and Bob. Gawron et al. [19] have studied noise effects in
quantum magic squares game. They have shown that the probability of success
can be used to determine the characteristics of quantum channels. Recently,
Sousa et al. [22] have proposed a quantum game which can be used to control
the access of processes to the CPU in a quantum computer. More recently,
James et al. [23] have analyzed the quantum penny flip game using geometric
algebra.

Decoherence is an integral part of the theory of quantum computation and
communication. A major problem of quantum communication is to faithfully
transmit unknown quantum states through a noisy quantum channel. When
quantum information is sent through a channel (optical fiber), the carriers
of the information interact with the channel and get entangled with its many
degrees of freedom. This gives rise to the phenomenon of decoherence on the
state space of the information carriers. To deal with the problem of
decoherence, two methods have been developed, known as quantum error
correction [24] and entanglement purification [25]. When quantum information
is processed in the real-world, the decoherence caused by the external
environment is inevitable. Implementation of decoherence in quantum games
have been studied by different authors [6, 26]. Recently, interest has been
developed in implementing quantum memory in the field of quantum game theory
[8].

Quantum channels are implemented by suitable quantum devices consisting of
intrinsic degrees of freedom associated with the environment and acting on
the system via particular interactions between the system and the
environment. The assumption that noise is uncorrelated between successive
uses of a channel is not realistic. Hence memory effects need to be taken
into account. Quantum channels with memory [27-29] are the natural
theoretical framework for the study of any noisy quantum communication
system where correlation time is longer than the time between consecutive
uses of the channel. A more general model of a quantum channel with memory
was introduced by Bowen and Mancini [30] and also studied by Kretschmann and
Werner [31].

In this paper, we study the quantum magic squares game influenced by
different memory channels, such as amplitude damping, depolarizing, phase
damping, bit-flip, phase-flip and bit-phase-flip channels, parameterized by
a quantum noise parameter $\alpha $\ and memory parameter $%
\mu
$. Here $\alpha \in \lbrack 0,1]$ and $\mu \in \lbrack 0,1]$ represent the
lower and upper limits of quantum noise parameter and memory parameter
respectively. It is seen that the behaviour of the damping (amplitude and
phase damping) channels is different as compared to the depolarizing and
flipping channels. Therefore, the quantum channels can easily be
distinguished using the success probability of the game.

\section{The magic squares game}

The magic squares game is a two-player game presented by Aravind [32] which
was primarily built on the work done by Mermin [33]. The magic square is a $%
3\times 3$ matrix filled with numbers $0$ or $1,$ such that the sum of
entries in each row is even and the sum of entries in each column is odd.
However, it is impossible to have such a matrix with all the rows having
even parity and all the columns having odd parity. Therefore, there can be
no classical strategy that always wins. In the magic squares game, there are
two players, say, Alice and Bob. Alice is given the number of the row and
Bob is given the number of the column. Alice gives the entries for a row and
Bob gives entries for a column. The winning condition is that the parity of
the row must be even, the parity of the column must be odd, and the
intersection of the given row and column must be same. During the game,
Alice and Bob are not allowed to communicate with each other. There exists a
(classical) strategy that leads to winning probability of $8/9$. If parties
are allowed to share a quantum state they can achieve probability $1$. An
interesting feature of this game is that it does not require a promise.

In the quantum version of this game [21], Alice and Bob share an entangled
quantum state of the form

\begin{equation}
|\Psi _{in}\rangle =\frac{1}{2}\left( |0011\rangle -|1100\rangle
-|0110\rangle +|1001\rangle \right)  \label{init}
\end{equation}%
In equation (\ref{init}) the first two qubits correspond to Alice, whereas
the last two qubits correspond to Bob. Depending upon their inputs (the
specific row and column to be filled in) Alice and Bob apply unitary
operators $A_{i}\otimes I$ and $I\otimes B_{j}$, respectively,%
\begin{eqnarray}
A_{1} &=&\frac{1}{\sqrt{2}}\left[
\begin{array}{cccc}
i & 0 & 0 & 1 \\
0 & -i & 1 & 0 \\
0 & i & 1 & 0 \\
1 & 0 & 0 & i%
\end{array}%
\right] ,\text{\qquad\ \ \ \ \ }B_{1}=\frac{1}{2}\left[
\begin{array}{cccc}
i & -i & 1 & 1 \\
-i & -i & 1 & -1 \\
1 & 1 & -i & i \\
-i & i & 1 & 1%
\end{array}%
\right]  \notag \\
A_{2} &=&\frac{1}{2}\left[
\begin{array}{cccc}
i & 1 & 1 & i \\
-i & 1 & -1 & i \\
i & 1 & -1 & -i \\
-i & 1 & 1 & -i%
\end{array}%
\right] ,\text{\qquad\ \ \ }B_{2}=\frac{1}{2}\left[
\begin{array}{cccc}
-1 & i & 1 & i \\
1 & i & 1 & -i \\
1 & -i & 1 & i \\
-1 & -i & 1 & -i%
\end{array}%
\right]  \notag \\
A_{3} &=&\frac{1}{2}\left[
\begin{array}{cccc}
-1 & -1 & -1 & 1 \\
1 & 1 & -1 & 1 \\
1 & -1 & 1 & 1 \\
1 & -1 & -1 & -1%
\end{array}%
\right] ,\text{\qquad }B_{3}=\frac{1}{\sqrt{2}}\left[
\begin{array}{cccc}
1 & 0 & 0 & 1 \\
-1 & 0 & 0 & 1 \\
0 & 1 & 1 & 0 \\
0 & 1 & -1 & 0%
\end{array}%
\right]
\end{eqnarray}%
where $i$ and $j$ denote the corresponding input. Finally, Alice and Bob
measure their qubits in the computational basis. Further steps to calculate
the success probability of the game are given in the next section.

\subsection{Quantum memory channels}

A major hurdle in the path of efficient information transmission is the
presence of noise, in both classical and quantum channels. This noise causes
a distortion of the information sent through the channel. Error correcting
codes are used to overcome this problem. Messages are encoded into
code-words, which are then sent through the channel. Information
transmission is said to be reliable if the probability of error, in decoding
the output of the channel, vanishes asymptotically in the number of uses of
the channel. A basic question of information theory is whether there is any
advantage in using entangled states as input states. That is, whether or not
encoding the classical data into entangled rather than separable states
increases the mutual information. For the case when multiple uses of the
channel are not correlated, there is no advantage in using entangled states.
Correlated noise, also referred as memory in the literature, acts on
consecutive uses of the channels. However, in general, one may want to
encode classical data into entangled strings or consecutive uses of the
channel may be correlated to each other. Hence, we are dealing with a
strongly correlated quantum system, the correlation of which results from
the memory of the channel itself.

In\ ref. [26] a Pauli channel with partial memory was studied. The action of
the channel on two consecutive qubits is given by the Kraus operators%
\begin{equation}
A_{ij}=\sqrt{\alpha _{i}[(1-\mu )\alpha _{j}+\mu \delta _{ij}]}\sigma
_{i}\otimes \sigma _{j}
\end{equation}%
where $\sigma _{i}$ ($\sigma _{j})$ are usual Pauli matrices, $\alpha _{i}$ (%
$\alpha _{j}$) represent the quantum noise and indices $i$ and $j$ runs from
$0$ to $3.$ The above expression means that with probability $\mu $ the
channel acts on the second qubit with the same error operator as on the
first qubit, and with probability $(1-\mu )$ it acts on the second qubit
independently. Physically the parameter $%
\mu
$ is determined by the relaxation time of the channel when a qubit passes
through it. In order to remove correlations, one can wait until the channel
has relaxed to its original state before sending the next qubit, however
this lowers the rate of information transfer. Thus it is necessary to
consider the performance of the channel for arbitrary values of $%
\mu
$ to reach a compromise between various factors which determine the final
rate of information transfer.\ Thus in passing through the channel any two
consecutive qubits undergo random independent (uncorrelated) errors with
probability ($1-%
\mu
)$ and identical (correlated) errors with probability $%
\mu
$. This should be the case if the channel has a memory depending on its
relaxation time and if we stream the qubits through it. The action of the
Pauli channels on $n$-qubits can be generalized in Kraus operator form as
given below%
\begin{equation}
A_{i_{1}.....i_{n}}=\sqrt{\alpha _{i_{n}}\prod\limits_{m=1}^{n-1}[(1-\mu
)\alpha _{i_{m}}+\mu \delta _{i_{m},i_{m+1}}]}\sigma _{i_{1}}\otimes
.....\otimes \sigma _{i_{n}}  \label{KO}
\end{equation}%
\ As stated above, with probability ($1-%
\mu
)$ the noise is uncorrelated and can be completely specified by the Kraus
operators
\begin{equation}
D_{ij}^{u}=\sqrt{\alpha _{i}\alpha _{j}}\sigma _{i}\otimes \sigma _{j},
\end{equation}%
and with probability $%
\mu
$ the noise is correlated (i.e. the channel has memory) which can be
specified by the Kraus operators
\begin{equation}
D_{kk}^{c}=\sqrt{\alpha _{k}}\sigma _{k}\otimes \sigma _{k},
\end{equation}%
A detailed list of single qubit Kraus operators for different quantum
channels with uncorrelated noise is given in table 1. The action of such a
channel if $n$ qubits are streamed through it, can be described in operator
sum representation as [2]
\begin{equation}
\rho _{f}=\sum\limits_{k_{1,}....,.k_{n}=0}^{n-1}(A_{k_{n}}\otimes
.....A_{k_{1}})\rho _{in}(A_{k_{1}}^{\dagger }\otimes
.....A_{k_{n}}^{\dagger })
\end{equation}%
where $\rho _{in}$ represents the initial density matrix for quantum state
and $A_{k_{n}}$\ are the Kraus operators expressed in equation (\ref{KO}).
The Kraus operators satisfy the completeness relation
\begin{equation}
\sum\limits_{k_{n}=0}^{n-1}A_{k_{n}}^{\dagger }A_{k_{n}}=1
\end{equation}%
However, the Kraus operators for a quantum amplitude damping channel with
correlated noise are given by Yeo and Skeen [27] as given as

\begin{equation}
A_{00}^{c}=\left[
\begin{array}{llll}
\cos \chi & 0 & 0 & 0 \\
0 & 1 & 0 & 0 \\
0 & 0 & 1 & 0 \\
0 & 0 & 0 & 1%
\end{array}%
\right] ,\ \ \ A_{11}^{c}=\left[
\begin{array}{llll}
0 & 0 & 0 & 0 \\
0 & 0 & 0 & 0 \\
0 & 0 & 0 & 0 \\
\sin \chi & 0 & 0 & 0%
\end{array}%
\right]
\end{equation}%
where, $0\leq \chi \leq \pi /2$ and is related to the quantum noise
parameter as
\begin{equation}
\sin \chi =\sqrt{\alpha }
\end{equation}%
It is clear that $A_{00}^{c}$ cannot be written as a tensor product of two
two-by-two matrices. This gives rise to the typical spooky action of the
channel: $\left\vert 01\right\rangle $ and $\left\vert 10\right\rangle $,
and any linear combination of them, and $\left\vert 11\right\rangle $ will
go through the channel undisturbed, but not $\left\vert 00\right\rangle .$%
The action of this non-unital channel is given by
\begin{equation}
\pi \rightarrow \rho =\Phi (\pi )=(1-\mu
)\sum\limits_{i,j=0}^{1}A_{ij}^{u}\pi A_{ij}^{u\dagger }+\mu
\sum\limits_{k=0}^{1}A_{kk}^{c}\pi A_{kk}^{c\dagger }
\end{equation}

The action of the super-operators provides a way of describing the evolution
of quantum states in a noisy environment. In our scheme, the Kraus operators
are of the dimension $2^{3}$. They are constructed from single qubit Kraus
operators by taking their tensor product over all $n^{3}$ combinations
\begin{equation}
A_{k}=\underset{k_{n}}{\otimes }A_{k_{n}}
\end{equation}%
where $n$ is the number of Kraus operator for a single qubit channel. The
final state of the game after the action of the channel can be obtained as
\begin{equation}
\rho _{f}=\Phi _{\alpha ,\mu }(\left\vert \Psi \right\rangle \left\langle
\Psi \right\vert )
\end{equation}%
where $\Phi _{\alpha ,\mu }$ is the super-operator realizing the quantum
channel parametrized by real numbers $\alpha $ and $\mu $. After the action
of the players unitary operations, the game's final state transforms to
\begin{equation}
\rho _{\acute{f}}=(A_{i}\otimes B_{j})(\Phi _{\alpha ,\mu }(\left\vert \Psi
\right\rangle \left\langle \Psi \right\vert )(A_{i}^{\dag }\otimes
B_{j}{}^{\dag })  \label{rf}
\end{equation}%
The probability of success $P_{i,j}(\alpha ,\mu )$ can be computed as the
probability of measuring $\rho _{\acute{f}}$ in the state indicating success%
\begin{equation}
P_{i,j}(\alpha ,\mu )=\text{Tr}(\rho _{\acute{f}}\sum\limits_{m}\left\vert
\xi _{m}\right\rangle \left\langle \xi _{m}\right\vert )
\end{equation}%
where $\left\vert \xi _{m}\right\rangle $\ are the states that imply success
and Tr represents the trace of the matrix. The mean probability of success
of the game is calculated from
\begin{equation}
\bar{P}_{i,j}(\alpha ,\mu )=\sum_{i,j\in \{1,2,3\}}P_{i,j}(\alpha ,\mu ).
\end{equation}%
It can be easily checked that the results of ref. \cite{Gawron} can be
reproduced if we put $\mu =0$\ in tables 2 and 3 for success probability and
mean success probability respectively.

\section{Results and discussions}

We compute the success probability $P_{i,j}(\alpha ,\mu )$ for different
inputs $(i,j\in \{1,2,3\})$ in the magic squares game for different quantum
channels such as depolarizing, amplitude damping, phase damping and flipping
channels. It is seen that the mean probability of success, $\bar{P}%
_{i,j}(\alpha ,\mu ),$ heavily depends on the quantum noise parameter $%
\alpha $ and memory parameter $\mu $ (see table 3). The game results for
each combination of operators $A_{i}$, $B_{j}$ for depolarizing, amplitude
damping, phase damping, bit flip, phase flip and bit-phase flip channels are
listed in table 2. In table 3, we present the results for the mean success
probability $\bar{P}_{i,j}(\alpha ,\mu )$. It is seen that in the case of
depolarizing channel, the success probability as the function of quantum
noise parameter $\alpha $ and memory parameter $\mu $ is the same for all
the possible inputs. For amplitude damping and phase damping channels, one
can observe three different types of functions. These functions are
increasing for these channels under the effect of memory. The depolarizing
function reaches its minimum for $\alpha =\mu =1/2$ and is symmetrical. It
is seen that in case of input $(1,3)$ the phase-flip channel does not
influence the probability of success. The same is true for input $(2,3)$ for
bit flip channel and also for input $(3,3)$ for bit-phase flip channel.
Hence, it is possible to distinguish these channels by looking at the
success probability of magic-squares game.

In figures 1 and 2, we plot mean success probability $\bar{P}(\alpha ,\mu )$
as a function of quantum noise parameter $\alpha $\ for memory parameter $%
\mu =0.5$\ and $\mu =1$\ respectively for amplitude damping, depolarizing,
phase damping and flipping channels. It is seen that the amplitude damping,
phase damping channels cause monotonic decrease of mean success probability
as a function of quantum noise parameter for $\mu <1$ (see figure 1). On the
other hand, depolarizing and flipping channels give symmetrical function.
However, for $\mu =1,$ the amplitude damping and phase damping channels
cause monotonic decrease of mean success probability. Where as in case of
depolarizing and flipping channels, the success functions attain their
maximum probability of success (see figure 2). In figures 3 and 4, we plot
mean success probability $\bar{P}(\alpha ,\mu )$ as a function of memory
parameter $\mu $ for $\alpha =0.5$\ and $\alpha =1$\ respectively, for
amplitude damping, depolarizing, phase damping and flipping channels. It is
seen that the mean success probability increases linearly as a function of
memory parameter $\mu ,$ for $\alpha <1$ (see figure 3). However, for
depolarizing and flipping channels it reaches its maximum at $\alpha =1$ for
the entire range of the memory parameter $\mu $.

In figures 5-8, we present the 3D graphs of mean success probability as a
function of $\alpha $\ and $\mu $\ for amplitude damping, depolarizing,
phase damping and flipping channels respectively. One can easily see that
the mean success probability is dependent on the quantum noise parameter $%
\alpha $ and memory parameter $\mu .$ It is seen that the mean success
probability decreases with $\alpha $\ and increases with $\mu ,$\ which
indicates that both the parameters influence the probability differently. It
is also seen that the behaviour of amplitude damping and phase damping
channels is same. On the other hand the behaviour of depolarizing channel is
similar to the flipping channels. Therefore, it is possible to distinguish
these channels by looking at the success probability of the game.

\section{Conclusions}

We study quantum magic squares game under the influence of quantum memory.
We analyze that how the quantum noise and memory of quantum channels can
influence the magic squares quantum pseudo-telepathy game. It is seen that
the mean success probability decreases with quantum noise parameter $\alpha $%
,\ for $\mu <1.$ Where as it increases with $\mu $\ for damping channels for
$\alpha <1$. On the other hand, for depolarizing and flipping channels the
success functions attain their maximum probability of success for entire
range of $\mu $. It is further seen that the behaviour of depolarizing and
flipping channels is rather different from the damping channels (amplitude
and phase damping). Therefore, we can conclude with the comment that, it is
possible to distinguish these channels by looking at the probability of
success of the game. In other words, the probability of success can be used
to distinguish the quantum channels.\newline

{\huge Figures captions}\newline
\textbf{Figure 1}. Mean success probability $\bar{P}(\alpha ,\mu )$ plotted
as a function of quantum noise parameter $\alpha $\ for memory parameter $%
\mu =0.5$\ for amplitude damping, depolarizing, phase damping and flipping
(phase flip, bit flip or bit-phase flip) channels.\newline
\textbf{Figure 2}. Mean success probability $\bar{P}(\alpha ,\mu )$ plotted
as a function of quantum noise parameter $\alpha $\ for memory parameter $%
\mu =1$\ for amplitude damping, depolarizing, phase damping and flipping
channels.\newline
\textbf{Figure 3}. Mean success probability $\bar{P}(\alpha ,\mu )$ plotted
as a function of memory parameter $\mu $\ for quantum noise parameter $%
\alpha =0.5$\ for amplitude damping, depolarizing, phase damping and
flipping channels.\newline
\textbf{Figure 4}. Mean success probability $\bar{P}(\alpha ,\mu )$\ plotted
as a function of memory parameter $\mu $\ for quantum noise parameter $%
\alpha =1$\ for amplitude damping, depolarizing, phase damping and flipping
channels.\newline
\textbf{Figure 5}. Mean success probability $\bar{P}(\alpha ,\mu )$\ plotted
as a function of memory parameter $\mu $\ and quantum noise parameter $%
\alpha $\ for amplitude damping channel.\newline
\textbf{Figure 6}. Mean success probability $\bar{P}(\alpha ,\mu )$\ plotted
as a function of memory parameter $\mu $\ and quantum noise parameter $%
\alpha $\ for depolarizing channel.\newline
\textbf{Figure 7}. Mean success probability $\bar{P}(\alpha ,\mu )$\ plotted
as a function of memory parameter $\mu $\ and quantum noise parameter $%
\alpha $\ for phase flip channel.\newline
\textbf{Figure 8}. Mean success probability $\bar{P}(\alpha ,\mu )$\ plotted
as a function of memory parameter $\mu $\ and quantum noise parameter $%
\alpha $\ for phase damping channel.\newline
{\Huge Tables Captions}\newline
\textbf{Table 1}. Single qubit Kraus operators for typical noise channels
such as depolarizing, amplitude damping, phase damping, phase flip, bit flip
and bit-phase flip where $\alpha $ represents the quantum noise parameter.%
\newline
\textbf{Table 2}. Success probability $P(\alpha ,\mu )$\ for all
combinations of magic squares game inputs for depolarizing, amplitude
damping, phase damping, phase flip, bit flip and bit-phase flip channels in
the presence of memory.\newline
\textbf{Table 3}. Mean success probability $\bar{P}(\alpha ,\mu )$ for
depolarizing, amplitude damping, phase damping and flipping (phase flip, bit
flip or bit-phase flip) channels in the presence of memory.\newpage

\begin{figure}[tbp]
\begin{center}
\vspace{-2cm} \includegraphics[scale=0.6]{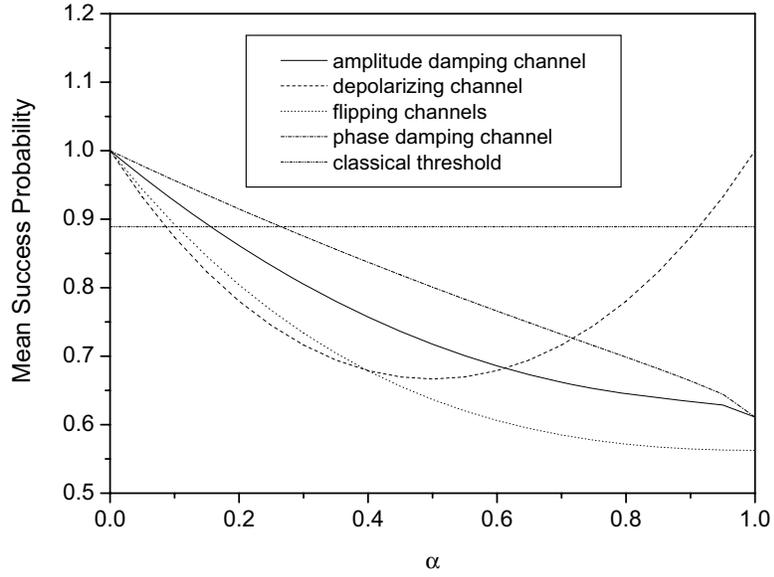} \\[0pt]
\end{center}
\caption{Mean success probability $\bar{P}(\protect\alpha ,\protect\mu )$
plotted as a function of quantum noise parameter $\protect\alpha $\ for
memory parameter $\protect\mu =0.5$\ for amplitude damping, depolarizing,
phase damping and flipping (phase flip, bit flip or bit-phase flip) channels.
}
\end{figure}
\begin{figure}[tbp]
\begin{center}
\vspace{-2cm} \includegraphics[scale=0.6]{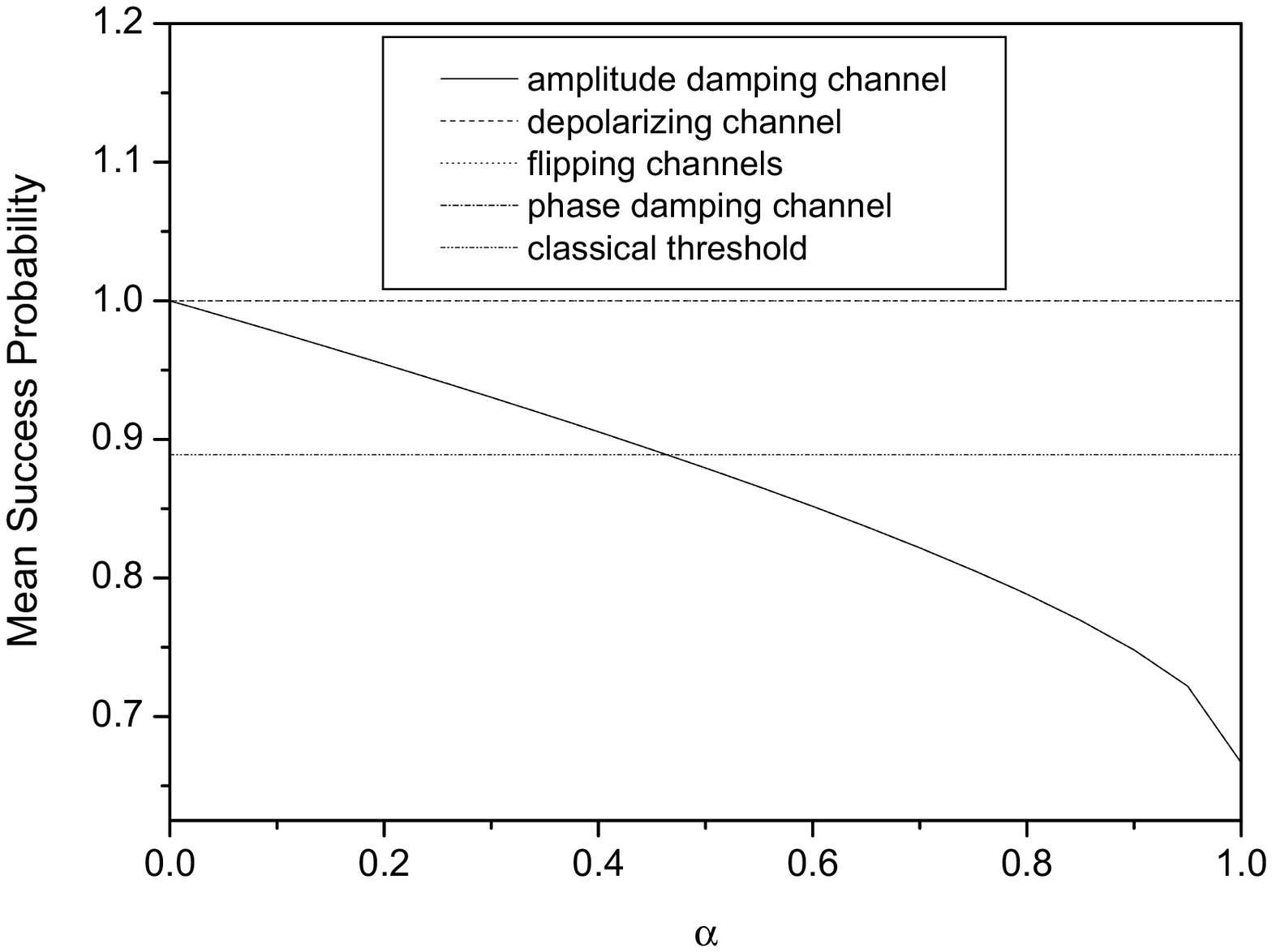} \\[0pt]
\end{center}
\caption{Mean success probability $\bar{P}(\protect\alpha ,\protect\mu )$
plotted as a function of quantum noise parameter $\protect\alpha $\ for
memory parameter $\protect\mu =1$\ for amplitude damping, depolarizing,
phase damping and flipping channels.}
\end{figure}
\begin{figure}[tbp]
\begin{center}
\vspace{-2cm} \includegraphics[scale=0.6]{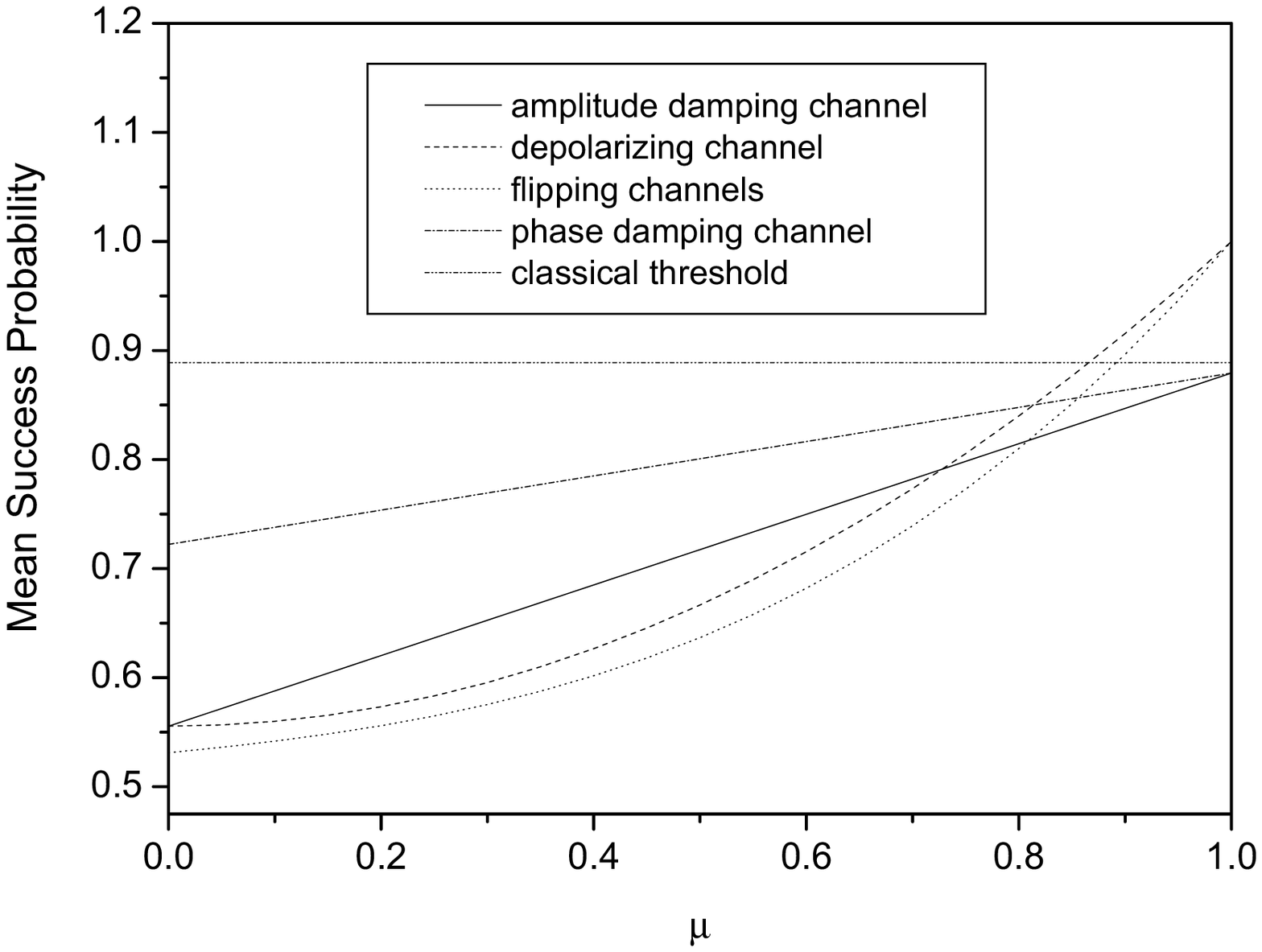} \\[0pt]
\end{center}
\caption{Mean success probability $\bar{P}(\protect\alpha ,\protect\mu )$
plotted as a function of memory parameter $\protect\mu $\ for quantum noise
parameter $\protect\alpha =0.5$\ for amplitude damping, depolarizing, phase
damping and flipping channels.}
\end{figure}
\begin{figure}[tbp]
\begin{center}
\vspace{-2cm} \includegraphics[scale=0.6]{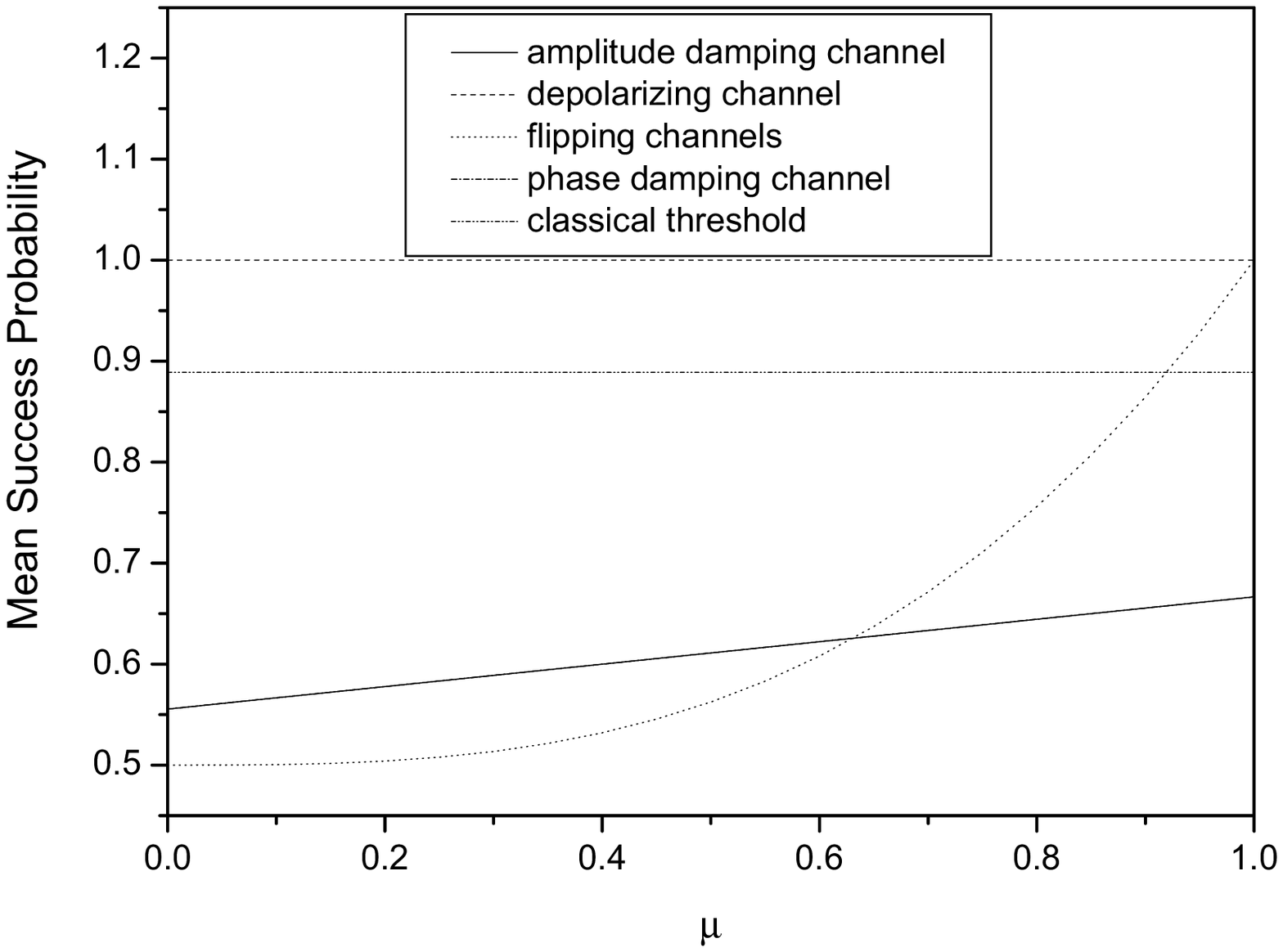} \\[0pt]
\end{center}
\caption{Mean success probability $\bar{P}(\protect\alpha ,\protect\mu )$\
plotted as a function of memory parameter $\protect\mu $\ for quantum noise
parameter $\protect\alpha =1$\ for amplitude damping, depolarizing, phase
damping and flipping channels.}
\end{figure}
\begin{figure}[tbp]
\begin{center}
\vspace{-2cm} \includegraphics[scale=0.6]{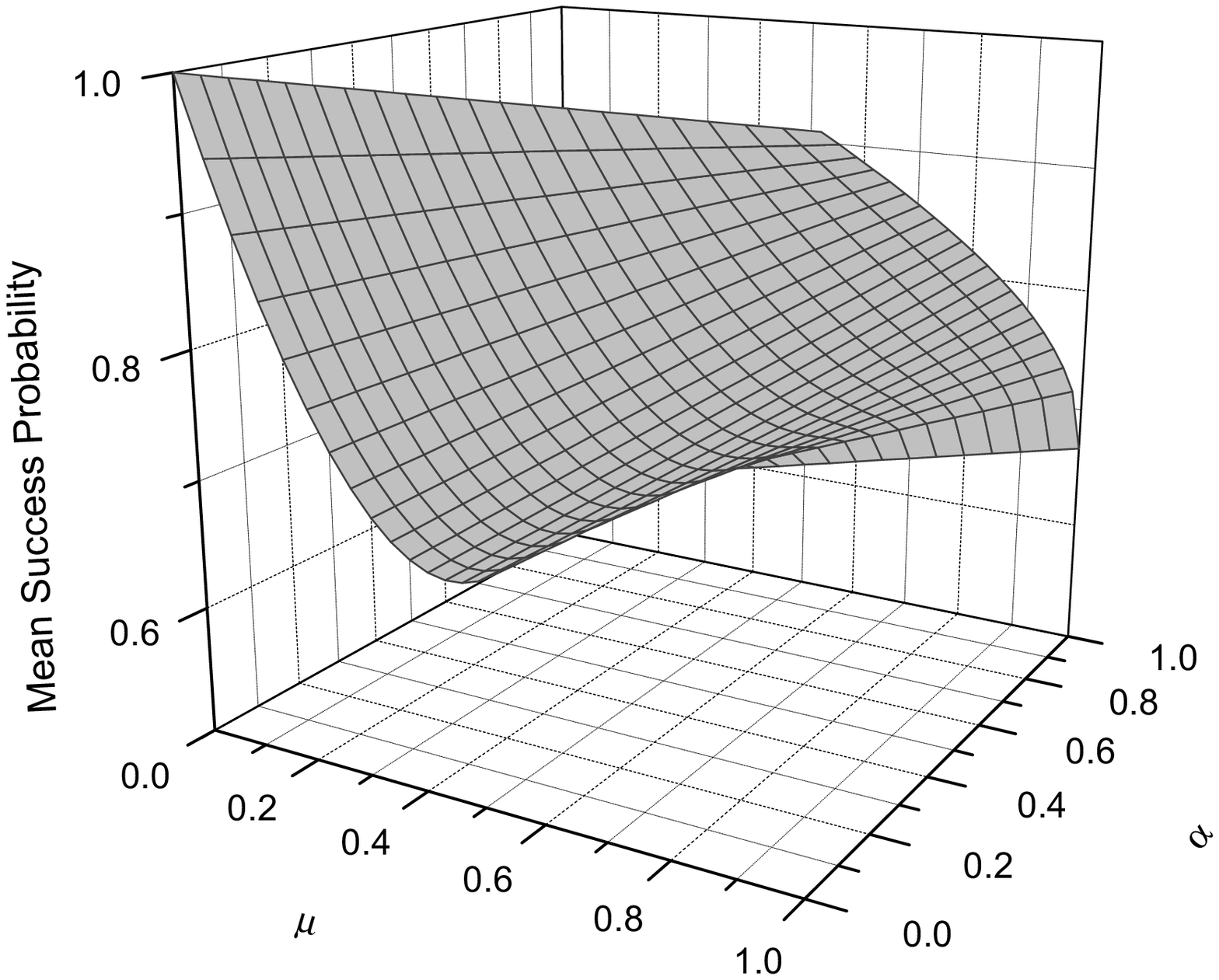} \\[0pt]
\end{center}
\caption{Mean success probability $\bar{P}(\protect\alpha ,\protect\mu )$\
plotted as a function of memory parameter $\protect\mu $\ and quantum noise
parameter $\protect\alpha $\ for amplitude damping channel.}
\end{figure}
\begin{figure}[tbp]
\begin{center}
\vspace{-2cm} \includegraphics[scale=0.6]{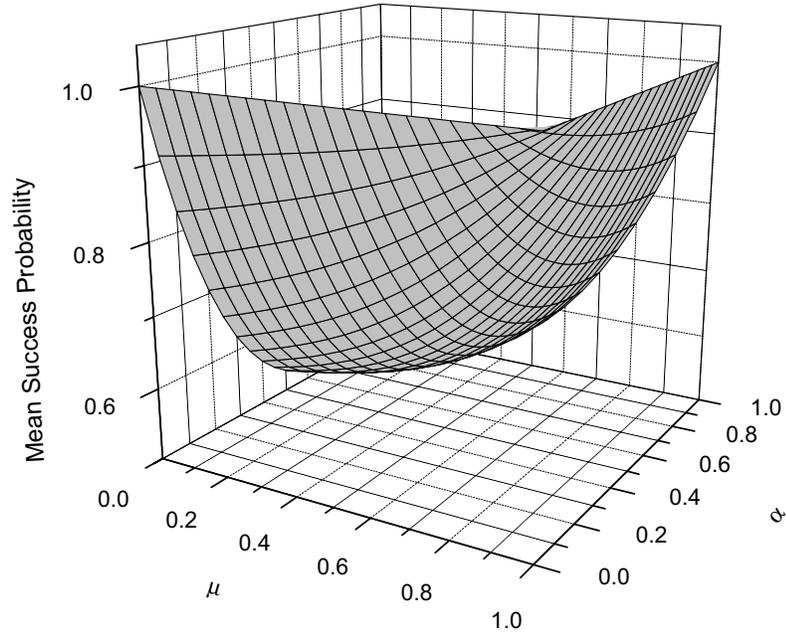} \\[0pt]
\end{center}
\caption{Mean success probability $\bar{P}(\protect\alpha ,\protect\mu )$\
plotted as a function of memory parameter $\protect\mu $\ and quantum noise
parameter $\protect\alpha $\ for depolarizing channel.}
\end{figure}
\begin{figure}[tbp]
\begin{center}
\vspace{-2cm} \includegraphics[scale=0.6]{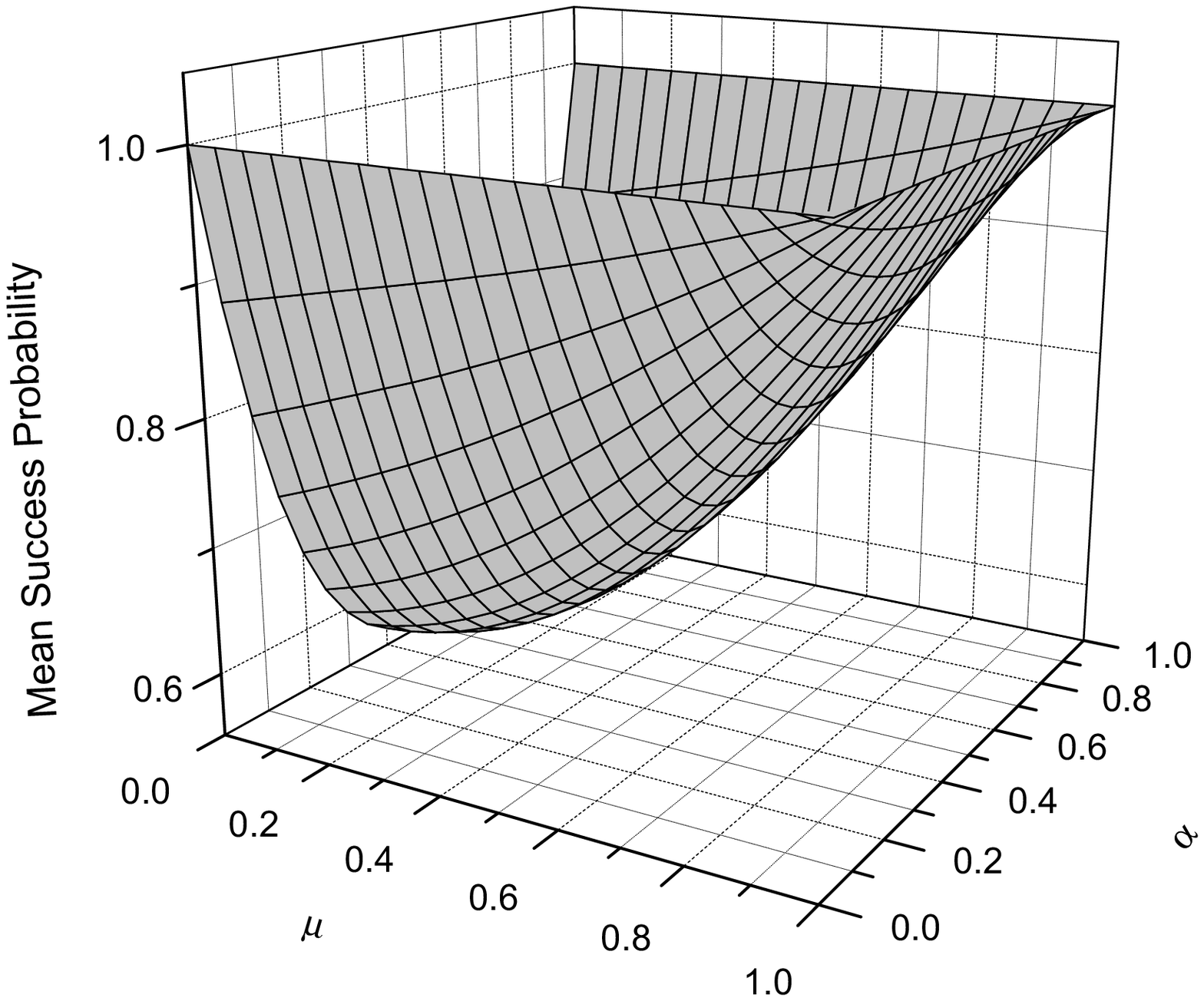} \\[0pt]
\end{center}
\caption{Mean success probability $\bar{P}(\protect\alpha ,\protect\mu )$\
plotted as a function of memory parameter $\protect\mu $\ and quantum noise
parameter $\protect\alpha $\ for phase flip channel.}
\end{figure}
\begin{figure}[tbp]
\begin{center}
\vspace{-2cm} \includegraphics[scale=0.6]{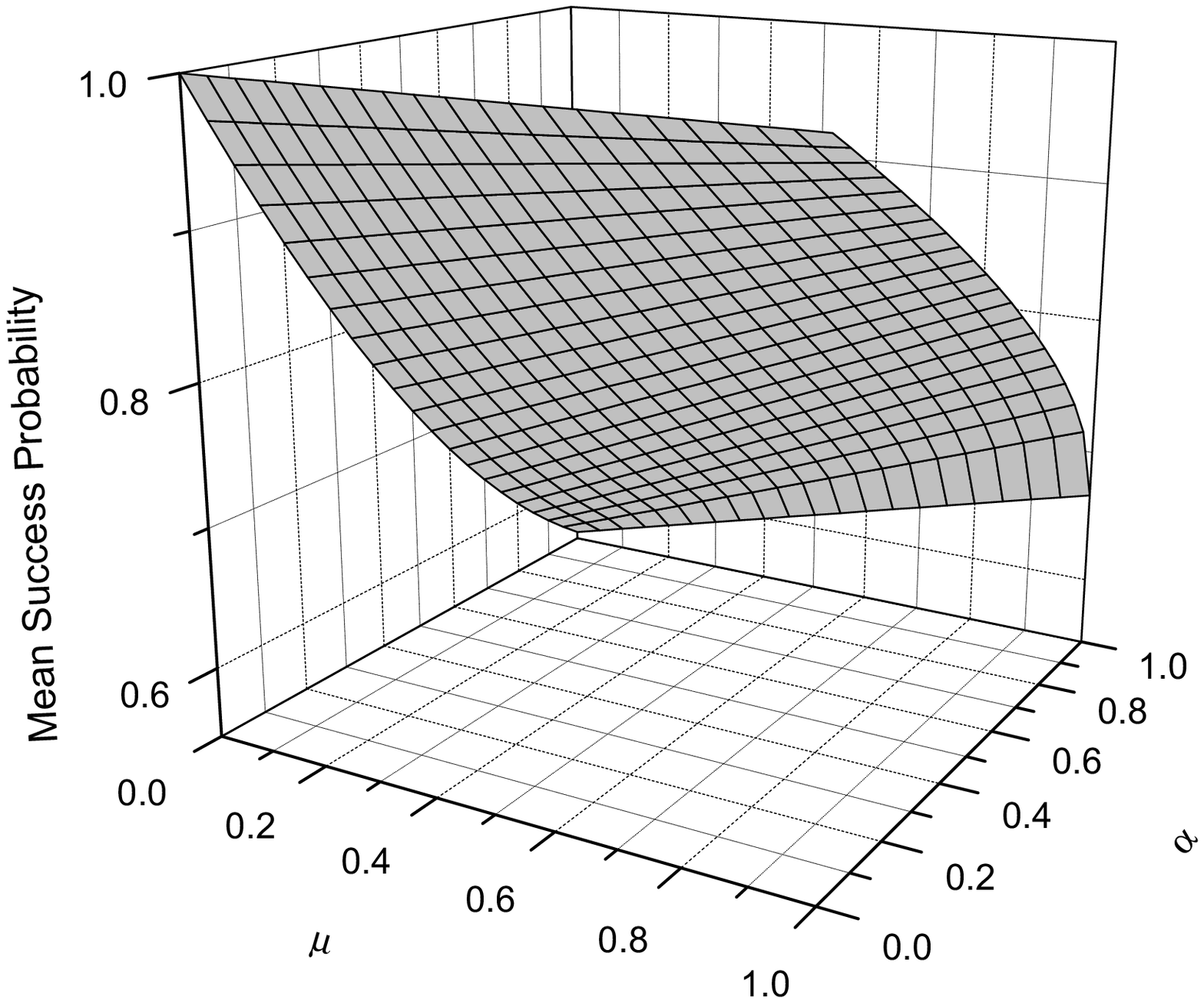} \\[0pt]
\end{center}
\caption{Mean success probability $\bar{P}(\protect\alpha ,\protect\mu )$\
plotted as a function of memory parameter $\protect\mu $\ and quantum noise
parameter $\protect\alpha $\ for phase damping channel.}
\end{figure}

\begin{table}[tbh]
\caption{Single qubit Kraus operators for typical noise channels such as
depolarizing, amplitude damping, phase damping, phase flip, bit flip and
bit-phase flip where $\protect\alpha $ represents the quantum noise
parameter.}$%
\begin{tabular}{|l|l|}
\hline
&  \\
$\text{Depolarizing channel}$ & $%
\begin{tabular}{l}
$E_{0}=\sqrt{1-\frac{3\alpha }{4}I},\quad E_{1}=\sqrt{\frac{\alpha }{4}}%
\sigma _{x}$ \\
$E_{2}=\sqrt{\frac{\alpha }{4}}\sigma _{y},\quad \quad $\ $\ E_{3}=\sqrt{%
\frac{\alpha }{4}}\sigma _{z}$%
\end{tabular}%
$ \\
&  \\ \hline
&  \\
$\text{Amplitude damping channel}$ & $E_{0}=\left[
\begin{array}{cc}
1 & 0 \\
0 & \sqrt{1-\alpha }%
\end{array}%
\right] ,$ $E_{1}=\left[
\begin{array}{cc}
0 & \sqrt{\alpha } \\
0 & 0%
\end{array}%
\right] $ \\
&  \\ \hline
&  \\
$\text{Phase damping channel}$ & $E_{0}=\left[
\begin{array}{cc}
1 & 0 \\
0 & \sqrt{1-\alpha }%
\end{array}%
\right] ,E_{1}=\left[
\begin{array}{cc}
0 & 0 \\
0 & \sqrt{\alpha }%
\end{array}%
\right] $ \\
&  \\ \hline
&  \\
$\text{Phase flip channel}$ & $E_{0}=\sqrt{1-\alpha }I,\quad E_{1}=\sqrt{%
\alpha }\sigma _{z}$ \\
&  \\ \hline
&  \\
$\text{Bit flip channel}$ & $E_{0}=\sqrt{1-\alpha }I,\quad E_{1}=\sqrt{%
\alpha }\sigma _{x}$ \\
&  \\ \hline
&  \\
$\text{Bit-phase flip channel}$ & $E_{0}=\sqrt{1-\alpha }I,\quad E_{1}=\sqrt{%
\alpha }\sigma _{y}$ \\
&  \\ \hline
\end{tabular}%
$%
\label{di-fit}
\end{table}
\newpage

\begin{table}[tbh]
\caption{Success probability $P(\protect\alpha ,\protect\mu )$\ for all
combinations of magic squares game inputs for depolarizing, amplitude
damping, phase damping, phase flip, bit flip and bit-phase flip channels in
the presence of memory.}%
\begin{tabular}{|l|l|}
\hline
\multicolumn{2}{|l|}{\ \ \ \ \ \ \ \ \ \ \ \ \ \ \ \ \ \ \ \ \ \ \ \ \ \ \ \
\ \ \ \ \ \ \ \ \ \ \ \ \ \ \ \ \ \ \ \ \ \ \ \ \ \ \ Depolarizing channel:}
\\ \hline
$\ \ \ \ \ \ \ \ \ \ \ \ \ \{\left( i,j\right) \mid i,j=1,2,3\}\ \ \ \ \ \ $
&
\begin{tabular}{l}
$P_{ij}(\alpha ,\mu )=\frac{1}{2}(2-4\alpha +3\mu \alpha +\mu ^{3}\alpha
+6\alpha ^{2}-9\mu \alpha ^{2}$ \\
$\qquad \qquad \quad +3\mu ^{2}\alpha ^{2}-4\alpha ^{3}+9\mu \alpha
^{3}-6\mu ^{2}\alpha ^{3}+\mu ^{3}\alpha ^{3}$ \\
$\qquad \qquad \quad +\alpha ^{4}-3\mu \alpha ^{4}+3\mu ^{2}\alpha ^{4}-\mu
^{3}\alpha ^{4})$%
\end{tabular}
\\ \hline
\multicolumn{2}{|l|}{\ \ \ \ \ \ \ \ \ \ \ \ \ \ \ \ \ \ \ \ \ \ \ \ \ \ \ \
\ \ \ \ \ \ \ \ \ \ \ \ \ \ \ \ \ \ \ \ \ \ \ \ \ \ \ Amplitude damping
channel:} \\ \hline
\begin{tabular}{l}
$i,j\in \{(1,1),(1,2),(2,3),(3,3)\}$ \\
$i,j\in \left. (1,3)\right. $ \\
$i,j\in \{(2,1),(2,2),(3,1),(3,2)\}$%
\end{tabular}
&
\begin{tabular}{l}
$P_{ij}(\alpha ,\mu )=\frac{1}{4}(4-4\alpha +3\mu \alpha +2\alpha ^{2}-2\mu
\alpha ^{2})$ \\
$P_{ij}(\alpha ,\mu )=1-2\alpha +2\mu \alpha +2\alpha ^{2}-2\mu \alpha ^{2}$
\\
$P_{ij}(\alpha ,\mu )=\frac{1}{2}(2-\mu +\mu \sqrt{1-\alpha }-3\alpha +3\mu
\alpha +2\alpha ^{2}-2\mu \alpha ^{2})$%
\end{tabular}
\\ \hline
\multicolumn{2}{|l|}{\ \ \ \ \ \ \ \ \ \ \ \ \ \ \ \ \ \ \ \ \ \ \ \ \ \ \ \
\ \ \ \ \ \ \ \ \ \ \ \ \ \ \ \ \ \ \ \ \ \ \ \ \ \ \ Phase damping channel:}
\\ \hline
\begin{tabular}{l}
$i,j\in \{(1,1),(1,2),(2,3),(3,3)\}$ \\
$i,j\in \left. (1,3)\right. $ \\
$i,j\in \{(2,1),(2,2),(3,1),(3,2)\}$%
\end{tabular}
&
\begin{tabular}{l}
$P_{ij}(\alpha ,\mu )=\frac{1}{4}(4-4\alpha +3\mu \alpha +2\alpha ^{2}-2\mu
\alpha ^{2})$ \\
$P_{ij}(\alpha ,\mu )=1$ \\
$P_{ij}(\alpha ,\mu )=\frac{1}{2}(2-\mu +\mu \sqrt{1-\alpha }-\alpha +\mu
\alpha )$%
\end{tabular}
\\ \hline
\multicolumn{2}{|l|}{\ \ \ \ \ \ \ \ \ \ \ \ \ \ \ \ \ \ \ \ \ \ \ \ \ \ \ \
\ \ \ \ \ \ \ \ \ \ \ \ \ \ \ \ \ \ \ \ \ \ \ \ \ \ \ Phase Flip channel:}
\\ \hline
\begin{tabular}{l}
$i,j\in \{(1,1),(1,2),(2,3),(3,3)\}$ \\
\\
\\
$i,j\in \left. (1,3)\right. $ \\
$i,j\in \{(2,1),(2,2),(3,1),(3,2)\}$%
\end{tabular}
&
\begin{tabular}{l}
$P_{ij}(\alpha ,\mu )=1-4\alpha +6\mu \alpha -4\mu ^{2}\alpha +2\mu
^{3}\alpha +12\alpha ^{2}+24\mu ^{2}\alpha ^{4}$ \\
$\qquad \qquad \quad -30\mu \alpha ^{2}+28\mu ^{2}\alpha ^{2}-10\mu
^{3}\alpha ^{2}-16\alpha ^{3}+48\mu \alpha ^{3}$ \\
$\qquad \qquad \quad -48\mu ^{2}\alpha ^{3}+16\mu ^{3}\alpha ^{3}+8\alpha
^{4}-24\mu \alpha ^{4}-8\mu ^{3}\alpha ^{4}$ \\
$P_{ij}(\alpha ,\mu )=1$ \\
$P_{ij}(\alpha ,\mu )=1-2\alpha +2\mu ^{2}\alpha +2\alpha ^{2}-2\mu
^{2}\alpha ^{2}$%
\end{tabular}
\\ \hline
\multicolumn{2}{|l|}{\ \ \ \ \ \ \ \ \ \ \ \ \ \ \ \ \ \ \ \ \ \ \ \ \ \ \ \
\ \ \ \ \ \ \ \ \ \ \ \ \ \ \ \ \ \ \ \ \ \ \ \ \ \ \ Bit flip channel:} \\
\hline
\begin{tabular}{l}
$i,j\in \{(1,1),(1,2),(3,1),(3,2)\}$ \\
$i,j\in \left. (2,3)\right. $ \\
$i,j\in \{(1,3),(2,1),(2,2),(3,3)\}$ \\
\\
\end{tabular}
&
\begin{tabular}{l}
$P_{ij}(\alpha ,\mu )=1-2\alpha +2\mu ^{2}\alpha +2\alpha ^{2}-2\mu
^{2}\alpha ^{2}$ \\
$P_{ij}(\alpha ,\mu )=1$ \\
$P_{ij}(\alpha ,\mu )=1-4\alpha +6\mu \alpha -4\mu ^{2}\alpha +2\mu
^{3}\alpha +12\alpha ^{2}+24\mu ^{2}\alpha ^{4}$ \\
$\qquad \qquad \quad -30\mu \alpha ^{2}+28\mu ^{2}\alpha ^{2}-10\mu
^{3}\alpha ^{2}-16\alpha ^{3}+48\mu \alpha ^{3}$ \\
$\qquad \qquad \quad -48\mu ^{2}\alpha ^{3}+16\mu ^{3}\alpha ^{3}+8\alpha
^{4}-24\mu \alpha ^{4}-8\mu ^{3}\alpha ^{4}$%
\end{tabular}
\\ \hline
\multicolumn{2}{|l|}{\ \ \ \ \ \ \ \ \ \ \ \ \ \ \ \ \ \ \ \ \ \ \ \ \ \ \ \
\ \ \ \ \ \ \ \ \ \ \ \ \ \ \ \ \ \ \ \ \ \ \ \ \ \ \ Bit-phase flip channel:
} \\ \hline
\begin{tabular}{l}
$i,j\in \{(1,1),(1,2),(2,1),(2,2)\}$ \\
$i,j\in \left. (3,3)\right. $ \\
$i,j\in \{(1,3),(2,3),(3,1),(3,2)\}$ \\
\\
\end{tabular}
&
\begin{tabular}{l}
$P_{ij}(\alpha ,\mu )=1-2\alpha +2\mu ^{2}\alpha +2\alpha ^{2}-2\mu
^{2}\alpha ^{2}$ \\
$P_{ij}(\alpha ,\mu )=1$ \\
$P_{ij}(\alpha ,\mu )=1-4\alpha +6\mu \alpha -4\mu ^{2}\alpha +2\mu
^{3}\alpha +12\alpha ^{2}+24\mu ^{2}\alpha ^{4}$ \\
$\qquad \qquad \quad -30\mu \alpha ^{2}+28\mu ^{2}\alpha ^{2}-10\mu
^{3}\alpha ^{2}-16\alpha ^{3}+48\mu \alpha ^{3}$ \\
$\qquad \qquad \quad -48\mu ^{2}\alpha ^{3}+16\mu ^{3}\alpha ^{3}+8\alpha
^{4}-24\mu \alpha ^{4}-8\mu ^{3}\alpha ^{4}$%
\end{tabular}
\\ \hline
\end{tabular}%
\end{table}
\newpage

\begin{table}[tbh]
\caption{Mean success probability $\bar{P}(\protect\alpha ,\protect\mu )$
for depolarizing, amplitude damping, phase damping and flipping (phase flip,
bit flip or bit-phase flip) channels in the presence of memory.}%
\begin{tabular}{|ll|}
\hline
\multicolumn{2}{|l|}{\ \ \ \ \ \ \ \ \ \ \ \ \ \ \ \ \ \ \ Depolarizing
channel:} \\ \hline
&  \\
\begin{tabular}{l}
$\bar{P}(\alpha ,\mu )=$ \\
\end{tabular}
&
\begin{tabular}{l}
$\frac{1}{2}(2-4\alpha +3\mu \alpha +\mu ^{3}\alpha +6\alpha ^{2}-9\mu
\alpha ^{2}+3\mu ^{2}\alpha ^{2}-4\alpha ^{3}$ \\
$+9\mu \alpha ^{3}-6\mu ^{2}\alpha ^{3}+\mu ^{3}\alpha ^{3}+\alpha ^{4}-3\mu
\alpha ^{4}+3\mu ^{2}\alpha ^{4}-\mu ^{3}\alpha ^{4})$%
\end{tabular}
\\
&  \\ \hline
\multicolumn{2}{|l|}{\ \ \ \ \ \ \ \ \ \ \ \ \ \ \ \ \ \ \ Amplitude damping
channel:} \\ \hline
&  \\
\begin{tabular}{l}
$\bar{P}(\alpha ,\mu )=$ \\
\end{tabular}
&
\begin{tabular}{l}
$1/9(9-2\mu +2\mu \sqrt{1-\alpha }-12\alpha +11\mu \alpha +8\alpha ^{2}-8\mu
\alpha ^{2})$ \\
\end{tabular}
\\
&  \\ \hline
\multicolumn{2}{|l|}{\ \ \ \ \ \ \ \ \ \ \ \ \ \ \ \ \ \ \ Phase damping
channel:} \\ \hline
&  \\
\begin{tabular}{l}
$\bar{P}(\alpha ,\mu )=$ \\
\end{tabular}
&
\begin{tabular}{l}
$1/9(9-2\mu +2\mu \sqrt{1-\alpha }-6\alpha +5\mu \alpha +2\alpha ^{2}-2\mu
\alpha ^{2})$ \\
\end{tabular}
\\
&  \\ \hline
\multicolumn{2}{|l|}{\ \ \ \ \ \ \ \ \ \ \ \ \ \ \ \ \ \ \ Flipping channels:
} \\ \hline
&  \\
\begin{tabular}{l}
$\bar{P}(\alpha ,\mu )=$ \\
\\
\\
\end{tabular}
&
\begin{tabular}{l}
$1/9(9-24\alpha +24\mu \alpha -8\mu ^{2}\alpha +8\mu ^{3}\alpha +56\alpha
^{2}-120\mu \alpha ^{2}$ \\
$+104\mu ^{2}\alpha ^{2}-40\mu ^{3}\alpha ^{2}-64\alpha ^{3}+192\mu \alpha
^{3}-192\mu ^{2}\alpha ^{3}$ \\
$+64\mu ^{3}\alpha ^{3}+32\alpha ^{4}-96\mu \alpha ^{4}+96\mu ^{2}\alpha
^{4}-32\mu ^{3}\alpha ^{4})$ \\
\end{tabular}
\\ \hline
\end{tabular}%
\end{table}

\end{document}